\documentclass[a4paper,12pt,dvips]{article}
\usepackage{graphicx}
\begin{document}
\begin{titlepage} 
\vspace{0.2in} 
\begin{center} {\LARGE \bf 
Covariant Formulation of the Invariant Measure for the 
Mixmaster Dynamics 
}
\vspace*{2cm}

{\bf Giovanni Imponente  \\
Giovanni Montani }\\ 

\vspace*{2cm}
ICRA---International Center for Relativistic Astrophysics \\ 
Phys. Dept., University of Rome `La Sapienza', \\Piazzale Aldo Moro 5, 00185 Roma, Italy\\
e-mail: imponente@icra.it, montani@icra.it
\vspace*{2cm}

\vspace{\stretch{1}}
\end{center} \indent
\begin{abstract}
{\bf {Abstract:}} 
We provide a Hamiltonian analysis of the Mixmaster Universe dynamics 
showing the covariant nature of its chaotic behavior with respect to any 
choice of time variable. We construct the appropriate invariant measure for the system 
(which relies on the existence of an ``energy-like'' constant of motion) 
without fixing 
the time gauge, i.e. the corresponding lapse function. 
The key point in our analysis consists of introducing generic 
Misner-Chitr\'e-like variables containing an arbitrary function, whose 
specification allows one to set up the same dynamical scheme in any time gauge.
\vspace{0.5cm}

PACS number(s): {04.20.Jb, 98.80.Dr}
\end{abstract}
\smallskip\

\end{titlepage}
The original Belinski-Khalatnikov-Lifshitz (BKL) analysis of the asymptotic 
dynamics of Bianchi type VIII and IX cosmological models \cite{BKL70} 
(the so-called Mixmaster universes \cite{M69}) showed the existence of a 
chaotic behavior in their approach to the initial cosmological singularity. 

Probably the most suitable description of this BKL oscillatory regime, 
especially in view of its chaotic nature, is via the Hamiltonian 
formulation of the Mixmaster Universe dynamics expressed in 
Misner-Chitr\'e-like variables \cite{C72},\cite{MTW}. 
The advantage of this approach, apart from the immediate 
interpretation of the system evolution as a chaotic scattering process, 
consists of replacing the discrete BKL map by a geodesic flow in a space of 
continuous variables. In this approach a key result is obtaining an invariant measure for the 
Mixmaster evolution \cite{CB83,KM97} (see also \cite{B82,M00}).   

In spite of the achievements made in describing the Mixmaster 
stochasticity, however, because of its relativistic nature, the
dynamics can be viewed from different reference frames, thus 
leaving open the question of the covariance of the observed chaos. 

Interest in these 
covariance aspects has increased in recent years in view of the 
contradictory and often dubious results that have emerged on this topic. 
The confusion which arises regarding the effect of a change of the time 
variable in this problem depends on some special properties of the Mixmaster 
model when represented as a dynamical system, in particular the vanishing 
of the Hamiltonian and its non-positive definite kinetic terms 
(a typical feature of a gravitational system). 
These special features 
prevent the direct application of the most common criteria provided by the 
theory of dynamical systems for characterizing chaotic behavior 
(for a review, see \cite{H94}). 

Although a whole line of research opened up, following this problem of 
covariant characterization for the chaos in the Mixmaster
model \cite{PU91}--\cite{SS94}, the first widely accepted indications 
in favor of covariance were derived 
by a fractal formalism in \cite{CL97} (see also \cite{ML00}).
Indeed the requirement of a covariant description 
of the Mixmaster chaoticity when viewed 
in terms of continuous dynamical variables, due to the discrete 
nature of the fractal approach, 
leaves this subtle question open and prevents a general 
consensus from being reached. 

Here we show how the derivation of an 
invariant measure for the Mixmaster model (performed in \cite{KM97,M00} 
within the framework of the statistical mechanics) can be extended to a generic time gauge 
(more directly than in previous approaches
relying on fractal methods \cite{CL97}) 
provided that suitable Misner-Chitr\'e-like 
variables are chosen. 
More precisely, after a standard Arnowitt-Deser-Misner (ADM) 
reduction of the variational problem when written in terms of a generic time 
variable, we show how asymptotically close to the cosmological 
singularity, the Mixmaster dynamics can be modeled by a two-dimensional 
point-universe randomizing in a closed (in the sense below precised) domain 
with fixed ``energy'' 
(just the ADM kinetic energy); since it is natural to represent such a 
system by a microcanonical ensemble, then the corresponding invariant 
measure is induced by the Liouville one. The key point, making our 
result reliable, is that we consider an approximation dynamically induced 
by the asymptotic vanishing of the metric determinant.   

As well-known \cite{MTW}, the Mixmaster dynamics is governed by the 
Lagrangian (written in the Misner variables \cite{M69})  \cite{MTW}
\begin{equation} 
L=\frac{6D}{N}\left[{-{\alpha}^{\prime}}^2+{{\beta_+}^{\prime}}^2+{{\beta_-}^{\prime}}^2\right]- 
\frac{N}{D}V\left(\alpha, \beta_+ , \beta_-\right). 
\label{a} 
\end{equation}  
Here ${()}^{\prime} \equiv \frac{d}{d\eta}()$, $D\equiv e^{3\alpha}$ 
and the potential $V\left(\alpha, \beta_+ , \beta_-\right)$ 
reads 
\begin{eqnarray} 
V&=&\frac{1}{2} \left( D^{4H_1}+D^{4H_2}+D^{4H_3}\right) + \nonumber \\
&-&D^{2H_1 +2H_2}\pm D^{2H_2 
+2H_3}\pm D^{2H_3 +2H_1} , 
\label{b} 
\end{eqnarray}  
where $(+)$ and $(-)$ refer respectively to Bianchi type VIII and IX, and 
the anisotropy parameters $H_i ~\left(i=1,2,3\right)$ are the functions \cite{KM97} 
\begin{eqnarray} 
H_1 &=& \frac{1}{3}+ \frac{\beta_+ + \sqrt{3} \beta_-}{3 \alpha} \nonumber \\ 
H_2 &=& \frac{1}{3}+ \frac{\beta_+ - \sqrt{3} \beta_-}{3 \alpha}   \\ 
H_3 &=&\frac{1}{3}- \frac{2\beta_+}{3 \alpha}  \nonumber . 
\label{c} 
\end{eqnarray} 
In the limit $D\rightarrow 0$ corresponding to approaching an initial singularity, the second set of three terms of the above potential 
are clearly negligible with respect to the first set of three terms (excluding the particular cases when two or three anisotropy parameters 
$H_i$ coincide). 

Next introduce generic Misner-Chitr\'e-like variables 
by the transformation 
\begin{eqnarray} 
\label{d}
\alpha &=& - e^{\Gamma \left(\tau\right)}\xi \nonumber \\ 
\beta_+ &=& e^{\Gamma \left(\tau\right)}\sqrt{\xi^2 -1}\cos \theta \\ 
\beta_- &=& e^{\Gamma \left(\tau\right)}\sqrt{\xi^2 -1}\sin \theta  ,\nonumber 
\end{eqnarray} 
where $1\leq\xi<\infty$, 
$0\leq\theta<2\pi$ and
$\Gamma(\tau)$ stands for a generic function of $\tau$ (taken to be identical to $\tau$ by Chitr\'e, who also sets
$\xi=\cosh{\zeta}$, $0\le\zeta <\infty$, see \cite{MTW}). 
In terms of these variables 
the Lagrangian (\ref{a}) becomes
\begin{equation} 
L=\frac{6 D}{N} \left[ \frac{{\left(e^{\Gamma } {\xi}^{\prime}\right)}^2}{\xi ^2 -1} +{\left(e^{\Gamma}
{\theta}^{\prime}\right)}^2\left(\xi ^2 -1\right) -{{\left(e^{\Gamma }\right)}^{\prime}}^2 \right] -\frac{N}{D}V \left( \Gamma \left(\tau\right), \xi, \theta \right) , 
\label{e} 
\end{equation} 
where $D$ is 
\begin{equation} 
D = \exp\left\{ -3 \xi e^{\Gamma \left(\tau\right)} \right\} .
\label{f} 
\end{equation} 
Hence, by the usual Legendre 
transformation we obtain the conjugate momenta 
\begin{eqnarray}
\label{io}
p_{\tau}&=& -\frac{12 D}{N}{\left(e^{\Gamma} \cdot \frac{d\Gamma}{d\tau}\right)}^2  {\tau}^{\prime}   \nonumber \\
p_{\xi}&=& \frac{12 D}{N}e^{2\Gamma} \frac{{\xi}^{\prime}}{{\xi}^2 -1}  \\
p_{\theta}&=& \frac{12 D}{N}e^{2\Gamma} {\theta}^{\prime}\left({\xi}^2 -1\right)  \nonumber 
\end{eqnarray}
and reformulating the dynamical problem in terms of the corresponding
Hamiltonian variational principle leads to
\begin{equation} 
\delta \int \left(   p_{\xi} {\xi}^{\prime} +  p_{\theta} {\theta}^{\prime}+
p_{\tau}  {\tau}^{\prime} - \frac{Ne^{-2\Gamma }}{24 D} {\cal H}
\right) d\eta =0 ,
\label{g} 
\end{equation}
where 
\begin{equation} 
{\cal H} = -\frac{{p_{\tau}}^2}{\left(\frac{d\Gamma }{d\tau}\right)^2} +  {p_{\xi}}^2\left(\xi ^2 -1\right)
+\frac{{p_{\theta}}^2}{\xi ^2 -1} +24 V e^{2\Gamma } . 
\label{h} 
\end{equation} 

\begin{figure}
\resizebox{\hsize}{!}{\includegraphics{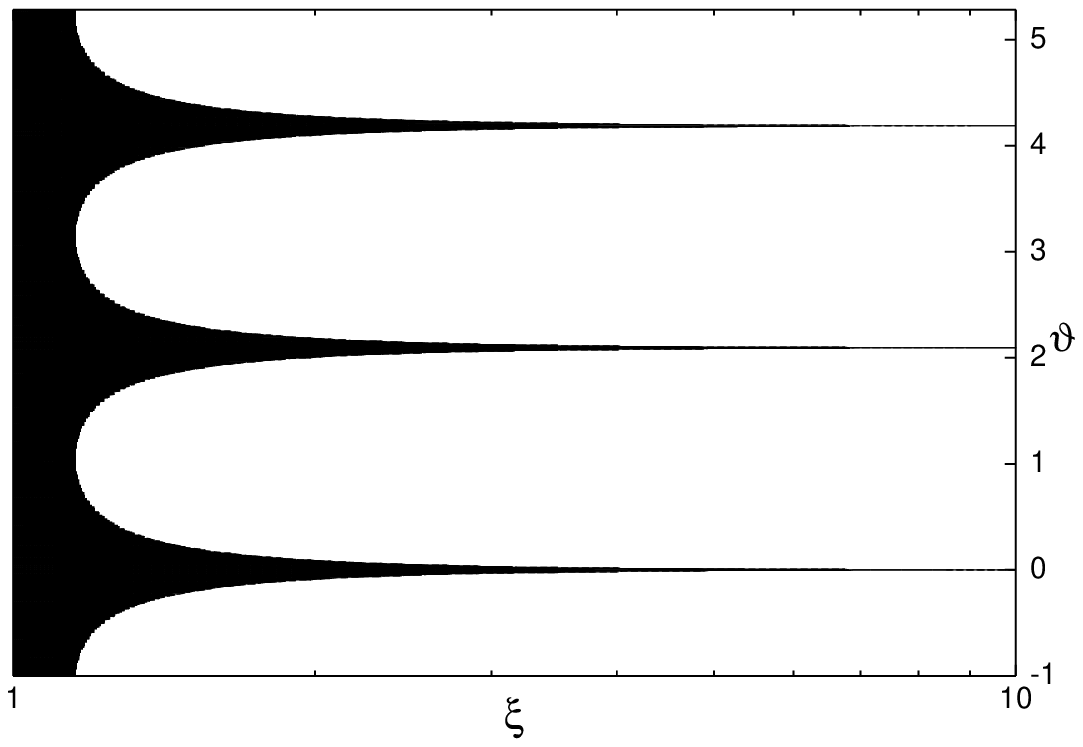}}
%
\caption{Asymptotic potential domain $\Gamma_H$ in the Lobachevsky plane $\theta,\xi$.}
\label{billiard}
\end{figure}

By varying (\ref{g}) with respect to $N$ we get the constraint ${\cal H}
=0$, whose solution is 
\begin{equation} 
- p_{\tau}\equiv \frac{d\Gamma }{d\tau} {\cal H}_{ADM} = \frac{d\Gamma }{d\tau}
\sqrt{\varepsilon ^2 +24 V e^{2\Gamma }} ,
\label{l} 
\end{equation}
where
\begin{equation}
\varepsilon ^2 \equiv  \left({\xi}^2 -1\right){p_{\xi}}^2 +\frac{{p_{\theta}}^2}{{\xi}^2 -1} .
\label{m} 
\end{equation} 
Thus we reduce the variational principle (\ref{g}) to the simpler one
\begin{equation} 
\delta \int \left(   p_{\xi} {\xi}^{\prime} +  p_{\theta} {\theta}^{\prime} 
- {\Gamma }^{\prime}{\cal H}_{ADM} \right) d\eta = 0; 
\label{ee} 
\end{equation} 
associated with this elimination, and from the first of (\ref{io}) and (\ref{l}) follows the time gauge relation 
\begin{equation} 
N\left(\eta\right) = \frac{12 D}{{\cal H}_{ADM}} e^{2\Gamma } \frac{d\Gamma }{d\tau} 
{\tau }^{\prime} . 
\label{o} 
\end{equation} 
Although now we choose the natural 
time gauge $\tau^{\prime }=1$ and rewrite the variational principle in terms of 
the time variable $\Gamma$ (indeed the Hamiltonian equations are equivalently viewed through the two 
time variables $\Gamma$ and $\tau$) 
leading to the 
variational principle 
\begin{equation} 
\delta \int \left(   p_{\xi} \frac{d\xi }{d\Gamma } +  p_{\theta} 
\frac{d\theta}{d\Gamma } 
- {\cal H}_{ADM} \right) d\Gamma = 0 , 
\label{p} 
\end{equation} 
nevertheless for any choice of time variable $\tau$ 
(i.e.\ $\tau=\eta$), there  
exists a corresponding function $\Gamma \left(\tau \right)$ 
(i.e. a set of Misner-Chitr\'e-like variables leading to the scheme (\ref{p})) 
defined by the (invertible) relation 
\begin{equation} 
\frac{d\Gamma }{d\tau} = 
\frac{{\cal H}_{ADM}}{12 D } N\left(\tau \right)e^{-2\Gamma } . 
\label{q} 
\end{equation} 
As a consequence of the variational principle (\ref{p}) we easily get the 
fundamental relation 
\begin{equation} 
\frac{\partial{\cal H}_{ADM}}{\partial \Gamma }=\frac{e^{2\Gamma }}
{ {\cal H}_{ADM}}
12\left( 2V+ \frac{\partial V}{\partial \Gamma } \right).
\label{r} 
\end{equation}
When expressed in terms of the Misner-Chitr\'e-like variables, the 
anisotropy parameters $H_i$ ($i = 1,2,3$) are independent of the 
variable $\Gamma$ 
\begin{eqnarray} 
\label{s} 
H_1 &=& \frac{1}{3} - \frac{\sqrt{\xi ^2 - 1}}{3\xi }\left(\cos\theta +
\sqrt{3}\sin\theta \right) \nonumber \\ 
H_2 &=& \frac{1}{3} - \frac{\sqrt{\xi ^2 - 1}}{3\xi }\left(\cos\theta -
\sqrt{3}\sin\theta \right)  \\ 
H_3 &=& \frac{1}{3} + 2\frac{\sqrt{\xi ^2 - 1}}{3\xi } \cos\theta ,\nonumber 
\end{eqnarray}
so as $D$ vanishes asymptotically approaching the initial singularity, the potential term 
can be modeled by the time-independent 
(see Fig.\ref{billiard}).
%
%
%
potential walls 
\begin{eqnarray}
\label{la}
U_\infty = \Theta _\infty \left(H_1\left(\xi, \theta\right)\right) &+& \Theta _\infty \left(H_2\left(\xi, \theta\right)\right) + \Theta _\infty \left(H_3\left(\xi, \theta\right)\right) \, , \\
\Theta _\infty \left(x\right) &=& + \infty \, \quad {\rm if} ~ x<0, 
\nonumber \\
\Theta _\infty \left(x\right) &=& 0 \, \quad \, \,\quad{\rm if} ~ x > 0 , \nonumber
\end{eqnarray}
and therefore in the region $\Pi _H$ where the potential vanishes,
we have by (\ref{r}) $d{\cal H}_{ADM}/d\Gamma = 0$, i.e. 
$\varepsilon = E = const$ (by other words the ADM Hamiltonian 
asymptotically approaches an integral of the motion). 

The invariant measure for this system can then be derived following 
the same approach presented in \cite{KM97,M00}. 

Indeed the Mixmaster dynamics in this approximation is equivalent to a two-dimensional point-universe moving within closed potential walls in a 
surface of constant negative curvature (the Lobachevsky plane \cite{KM97}), described by the line element
\begin{equation}
dl^2=E^2\left[ \left(\xi^2-1\right)d\xi^2 +\frac{d\theta^2}{\xi^2-1}\right], 
\label{uu}
\end{equation}
where $E$ is a constant.
From a geometrical point of view, the domain defined by the potential walls is 
not strictly closed, since there are three directions corresponding to the three corners in the
traditional ADM-Misner picture from which the point-universe could in principle escape (see Fig.\ref{billiard}).

However, as well known, for the Bianchi models under consideration, the only case in which there exists an 
asymptotic solution of the field equations with this behavior corresponds to two scale factors being equal
to each other; but as shown in \cite{BK69}, these cases are dynamically unstable and therefore correspond 
to sets of zero measure in the space of the initial conditions. For this reason it does not make sense 
to speak of a probability to reach certain configurations and the domain is {\it de facto} dynamically closed.

The bounces (billiard configuration) against 
the potential walls together with the instability of the geodesic flow on a 
closed domain of the Lobachevsky plane cause the point-universe dynamics to have a stochastic feature.
Indeed just the condition that the potential-free motion takes place in the Lobachevsky plane is itself not sufficient to infer the chaoticity of the dynamics, which is true for any Bianchi type model. It is the
compactness of the domain bounded by the potential walls that guarantees that the geodesic instability is transformed 
into a real stochastic behavior.
On the other hand, the possibility to speak of a stochastic scattering is justified
by the constant negative curvature of the Lobachevsky plane and therefore
these two notions (compactness and curvature) are both
necessary for our considerations. 
 
From the statistical mechanics point of view, such a system 
admits an ``energy-like'' constant of motion which corresponds
to the kinetic part of the ADM Hamiltonian $\varepsilon = E$. 
Since the point-universe randomizes within the phase-space associated 
with the closed domain $\Pi_H$, its stochastic dynamics admits a 
suitable ensemble representation which, in view of the existence of the ``energy-like''
constant of motion, has to have the natural feature of a {\it microcanonical} one.
Therefore the stochasticity of this system is appropriately described in terms of the
Liouville invariant measure
\begin{equation} 
d\varrho = const\, \delta \left(E - \varepsilon \right)d\xi d\theta dp_{\xi }dp_{\theta }  
\label{u} 
\end{equation} 
where $\delta\left(x\right)$ denotes the Dirac function, 
characterizing the {\it microcanonical ensemble}.

It is important to note that the particular value taken by the constant 
$\varepsilon$ $(\varepsilon=E)$ cannot influence 
the stochastic property of the system
and must be fixed by the initial conditions. 
To remove this useless information from the statistical dynamics under consideration,  
we must integrate over all admissible values of $\varepsilon$. To do this 
it is convenient to introduce the natural variables 
$(\varepsilon,\phi)$ in place of $(p_\xi,p_\theta)$ by
\begin{eqnarray} 
p_{\xi } &=& \frac{\varepsilon}{\sqrt{\xi ^2 - 1}}\cos\phi \nonumber \\ 
p_{\theta } &=& \varepsilon \sqrt{\xi ^2 - 1}\sin\phi ,  
\label{v} 
\end{eqnarray} 
where $0 \leq \phi < 2\pi$.
Integrating over all 
possible values of $\varepsilon$ removes the Dirac delta function, 
leading to the uniform (normalized) invariant measure 
\begin{equation} 
d\mu = d\xi d\theta d\phi \frac{1}{8\pi ^2} . 
\label{x} 
\end{equation} 
It is important to emphasize that the approximation on which our 
analysis is based (i.e. the potential wall model) is reliable since it is
dynamically induced no matter what time variable $\tau$ is used. 

The analysis developed in \cite{M00} shows that it is possible to 
derive asymptotically the Liouville theorem for the reduced 
phase space $\{\xi, \theta, \phi\}$. The relative continuity equation for the 
probability distribution $w\left(\Gamma, \theta, \phi\right)$ can be expressed as
\begin{eqnarray}
\label{xx}
\frac{\partial w}{\partial \Gamma} 
&+&\frac{d \xi}{d\Gamma} \frac{\partial w}{\partial\xi} 
+\frac{d \theta}{d\Gamma} \frac{\partial w}{\partial\theta}
+\frac{d \phi}{d\Gamma} \frac{\partial w}{\partial\phi} = \nonumber \\
&=& \frac{\partial w}{\partial \Gamma} +\sqrt{\xi^2-1}\cos\phi \frac{\partial w}{\partial\xi} + \nonumber \\
&\, +&\frac{\sin\phi}{\sqrt{\xi^2-1}}\frac{\partial w}{\partial\theta} 
-\frac{\xi\sin\phi}{\sqrt{\xi^2-1}}  \frac{\partial w}{\partial\phi} =0 \, .
\end{eqnarray}

One can immediately check that the invariant measure (\ref{x}) is a stationary solution of this continuity equation.

In fact, by applying 
the Landau-Raichoudhury theorem\footnote{This theorem, 
based on the mathematical assumptions underlying the dynamics of the Einstein equations, states that 
in a synchronous reference frame there always exists a value of the time at which the metric determinant vanishes and that in this time variable the zero is approached monotonically \cite{LF}.} 
near the initial singularity (which occurs by convention at $T=0$, 
where $T$ denotes the synchronous time, i.e. 
$dT=- N\left(\tau\right)d\tau$), we easily obtain the result 
that $D$ vanishes monotonically (i.e. for $T\rightarrow 0$ we 
have $d \ln D/d T > 0$). 
In terms of the adopted time variable $\tau$ 
$\left(D\rightarrow 0 \Rightarrow \Gamma\left(\tau\right) \rightarrow \infty\right)$, we have
\begin{equation} 
\frac{d \ln D}{d \tau } = 
\frac{d \ln D}{d T}\frac{dT}{d\tau } = 
- \frac{d \ln D}{d T}N\left(\tau \right)  
\label{x1} 
\end{equation}
and therefore $D$ vanishes monotonically for increasing $\tau$ as 
soon as $d\Gamma / d\tau >0$ according to (\ref{q}). 

The key point of our analysis is that any stationary solution 
of the Liouville theorem\footnote{When using a generic time variable $\tau$, 
the right-hand side of equation \ref{xx} in general is no longer (asymptotically) vanishing, 
but yet negligible with respect to the left-hand one.}, 
like (\ref{u}), remains valid 
for any choice of the time variable $\tau$; indeed the 
construction presented in \cite{M00} 
of the (restricted) Liouville theorem (\ref{xx}) 
shows the existence of such 
properties even for the invariant measure (\ref{x}).
Clearly the knowledge of the invariant measure (\ref{x}) provides a 
satisfactory statistical representation of the system for any choice of time variable,
since it allows one to calculate the asymptotic average 
values (as well as higher order moments) of any dynamical variable involved in the problem.

We conclude by remarking that when approaching the singularity 
$\Gamma\left(\tau\right) \rightarrow \infty$ (i.e. ${\cal H}_{ADM}\rightarrow E$), 
the time gauge relation (\ref{q}) simplifies to
\begin{equation} 
\label{qq}
\frac{d\Gamma }{d\tau} = 
N\left(\tau \right) \frac{E e^{-2\Gamma +3\xi e^{\Gamma}}}{12 }e^{-2\Gamma } . 
\label{x2} 
\end{equation} 
According to the analysis presented in \cite{M00}, the asymptotic functions 
$\xi\left(\Gamma\right), \theta\left(\Gamma\right), \phi\left(\Gamma\right)$ during free geodesic motion are governed by the equations
\begin{eqnarray}
\frac{d\xi}{d\Gamma}&=&\sqrt{\xi^2-1}\cos\phi \nonumber \\
\frac{d\theta}{d\Gamma}&=&\frac{\sin\phi}{\sqrt{\xi^2-1}} \\
\frac{d\phi}{d\Gamma}&=&-\frac{\xi\sin\phi}{\sqrt{\xi^2-1}} .\nonumber 
\end{eqnarray}
Once the solution $\xi\left(\Gamma\right)$ is obtained in the parametric form
\begin{eqnarray}
\label{xx1}
\xi \left(\phi\right)&=&\frac{\rho}{{\sin^2\phi}} \nonumber \\
\Gamma \left( \phi \right)&=&-a \left[ 
- \frac{1}{2} \frac{\rho \cos \phi ~{\rm arctanh} \left(\frac{1}{2} \frac{{\rho^2} +a^2 {\cos^2\phi}}{a\rho \cos \phi}\right)}{a\rho \cos \phi} \right] +b \nonumber \\
\rho&\equiv&\sqrt{a^2 +\sin^2\phi} \, \quad a,b=const.\in \Re 
\end{eqnarray}
equation (\ref{qq}) for free geodesic motion  reduces to a simple
differential equation for the function $\Gamma\left(\tau\right)$. 

However, as shown by our analysis, the global behavior of $\xi$ along the whole geodesic flow is described
by the invariant measure (\ref{x}) and therefore relation (\ref{qq}) indeed takes  a stochastic character. 
In other words, if we assign one of the two functions $\Gamma \left(\tau\right)$ or $N\left(\tau\right)$ 
to an arbitrary analytic functional form, then the other one will exhibit stochastic behavior.
Finally, by retaining the same dynamical scheme adopted to construct the 
invariant measure, we see how the one-to-one correspondence between any lapse function $N\left(\tau\right)$ and the associated 
set of Misner-Chit\`e-like variables (\ref{d}) guarantees the covariant nature with respect to the time-gauge of the Mixmaster universe stochastic behavior, when viewed in the framework of statistical mechanics.

Of course, our analysis leaves open the question concerning the covariance of the Mixmaster chaos with respect to the choice of configurational variables which are not  Misner-Chitr\'e-like.

\vspace{0.5cm} 

\section*{Acknowledgments}
We are very grateful to Vladimir Belinski and Remo Ruffini for their valuable comments on this 
subject. Robert Jantzen and Carlo Bianco are thanked respectively for the help in rewriting this 
manuscript and in producing the figure.

\small

%
%
\vspace{5cm}

\end{document}